\documentclass[aip,amsmath,amssymb,reprint]{revtex4-1}
\usepackage{graphicx}
\usepackage{dcolumn}
\usepackage{xcolor}
\usepackage{bm}
\usepackage{enumerate}
\usepackage{cancel}
\usepackage[normalem]{ulem}
\usepackage{footmisc}
\usepackage[utf8]{inputenc}
\usepackage[T1]{fontenc}
\usepackage{mathptmx}
\usepackage{braket}
\usepackage{xr}
\makeatletter

\def\br{{\bf r}}
\def\bp{{\bf r}'}
\def\dd{{\rm d}}

\newcommand*{\addFileDependency}[1]{
  \typeout{(#1)}
  \@addtofilelist{#1}
  \IfFileExists{#1}{}{\typeout{No file #1.}}
}
\makeatother

    \addFileDependency{SI.aux}

\begin{document}

\preprint{AIP/123-QED}

\title{Efficient treatment of molecular excitations in the liquid phase environment via stochastic many-body theory}

\author{Guorong Weng}
\author{Vojt\v{e}ch Vl\v{c}ek}
\email{vlcek@ucsb.edu}
\affiliation{Department of Chemistry and Biochemistry, University of California, Santa Barbara, CA 93106-9510, U.S.A.}

\date{\today}

\begin{abstract}
Accurate predictions of charge excitation energies of molecules in the disordered condensed phase are central to the chemical reactivity, stability, and optoelectronic properties of molecules and critically depend on the specific environment. Herein, we develop a stochastic GW method for calculating these charge excitation energies. The approach employs maximally localized electronic states to define the electronic subspace of a molecule and the rest of the system, both of which are randomly sampled. We test the method on three solute-solvent systems: phenol, thymine, and phenylalanine in water. The results are in excellent agreement with the previous high-level calculations and available experimental data. The stochastic calculations for supercells representing the solvated systems are inexpensive and require $\sim 1000$ CPU$\cdot$hrs. We find that the coupling with the environment accounts for $\sim40\%$ of the total correlation energy. The solvent-to-solute feedback mechanism incorporated in the molecular correlation term causes up to 0.6 eV destabilization of the QP energy. Simulated photo-emission spectra exhibit red shifts, state-degeneracy lifting, and lifetime shortening. Our method provides an efficient approach for an accurate study of excitations of large molecules in realistic condensed phase environments.
\end{abstract}

\maketitle

\section{Introduction}

The charged excited states of molecules in the condensed phase are critical for understanding the molecular electronic structure under realistic conditions. Experimentally, developments in photo-emission spectroscopy (PES) have enabled direct measurement of ionization in the  liquid\cite{Winter2006,Slavicek2009,Seidel2011,Pluharova2013,Tentscher2015,Riley2018,Suzuki2019} or the solid phase\cite{Seki1989,Seki1996,Seki1997,Hill1998,Sugiyama1998,DAndrade2005,Seki2006}. PES characterizes individual quasiparticles (QP), i.e., holes and electrons that are ``dressed'' by interactions with their surroundings and have finite (excitation) lifetimes. In principle, the spectra thus provide a direct route to estimate how the environment affects the molecular electronic structure. However, understanding the role of the environmental couplings requires PES with a high energetic and spatio-temporal resolution to differentiate the origin of individual QPs. In practice, the theory thus remains the primary route to uncover the details of the couplings.

Multiple approaches have been proposed for the computational treatment of molecules in realistic environments. The methodologies range from polarizable continuum models (PCM)\cite{Tomasi1994,Cramer1999,Orozco2000,Tomasi2005} to quantum embedding methods.\cite{Gordon2012,Pruitt2014,Chibani2016,Sun2016,Jones2020,Toth2020,Ma2021} In either of these, the condensed phase is not captured uniformly: the (small) system of interest is typically treated by a high-level quantum mechanical (QM) theory, and the remainder, i.e., the surrounding environment, is represented by an affordable and approximate approach, e.g., classical molecular mechanics.  Embedding of the density functional theory (DFT)\cite{Sun2016jctc,Sun2016jpcc,Kronik2018,Bhandari2018,Boruah2020} is computationally inexpensive and yields eigenvalues corresponding to valence states, but these levels fundamentally do not correspond to QP energies and also contain no information about excited-state lifetimes.\footnote{The eigenvalues of the Kohn-Sham states are stationary solutions with infinite lifetimes and they, by construction, do not correspond to the QPs.} The state-of-the-art quantum chemistry approaches, e.g., CCSD(T)\cite{Winter2005,Dedikova2011} and EOM-CC\cite{Ghosh2011,Ghosh2012,Bose2016,Ren2017,Ghosh2017,Chakraborty2017,Sadybekov2017,Barrozo2019} provide accurate predictions for IP and EA. However, they do not capture QP lifetimes either and suffer from a high computational cost.  Hence, these methods are applied to either small clusters or within an embedding scheme. Matching of multiple approaches is non-trivial and may lead to artifacts, e.g., blue shifts of the ionization energies rather than red shifts.\cite{Cauet2010}

Green's function (GF) techniques\cite{martin_reining_ceperley_2016} are becoming more widespread in chemistry and provide an appealing alternative to other methods. In principle, the GF formalism fully captures QPs and their lifetimes, and it is widely applied to condensed matter problems.\cite{Quek2006,Tamblyn2011,Kharche2014,Li2016,Kang2016,Gaiduk2016,Rangel2016,Li2016PRB,Pham2017,Blumenthal2017,Li2018,Gaiduk2018,Vlcek2018PRM,Liu2019,Brooks2020,Weng2020} Further, the methodology is open for systematic improvements; embedding within the GF framework\cite{Chibani2016,Li2016,Kang2016,martin_reining_ceperley_2016,Li2018,Romanova2020} is conceptually straightforward and ``seamless.''  Finally, recent developments of stochastic GF methods\cite{Neuhauser2014,Vlcek2017,Vlcek2018,Vlcek2019,Romanova2020} enabled calculations for giant systems with thousands of electrons, practically treating large and inhomogeneous systems on equal footing.

In this work, we develop and implement a new computational technique that employs the stochastic approach to many-body calculations combined with the decomposition of the electron-electron interaction terms. We demonstrate the first-principles QM approach on a set of molecules in the liquid water environment. The method is applied to the \textit{entire} condensed phase; the description of the electronic states in the solute and the solvent is thus consistent. We analyze the dynamical couplings and show that the interactions between the molecule and environment are sensitive to the local molecular geometry and orientation. The non-local correlations are responsible for state reordering; the feedback from the solvent leads to significant changes in the simulated PES spectra of the solutes compared to their isolated counterparts. The methodology captures the red shifts in energy, lifting electronic state-degeneracy, and significant QP lifetime shortening. The results are in excellent agreement with experimental data, and the approach thus provides a unique tool to address the electronic properties in realistic environments.

\section{Methodologies}

\subsection{Quasiparticle and decomposition of the self-energy}\label{sec:QPandSE}

The GF formalism conveniently describes the dynamics of the QP excitation and directly yields experimentally accessible observables.\cite{Golze2019} In practice, the excitation energies and lifetimes are identified from the \textit{spectral function} $A(\omega) = {\rm Im} G(\omega)$, where $G(\omega)$ is the GF (representing the time-ordered product of electron creation and annihilation operators). Conceptually, the GF represents the probability amplitude associated with addition or removal at two distinct space-time points; hence it captures the \textit{propagation} of the excitation through the system. 

The maxima in $A(\omega)$, i.e., the excitation energies, correspond to the poles of $G(\omega)$ (on real axis). The QP propagator $G$ is ``dressed'' by many-body interactions and it is related to a (mean-field) non-interacting GF, $G_0$, via the Dyson equation: $G^{-1} = G_0^{-1} - \Sigma$, where the self-energy, $\Sigma$, emerges as the the central quantity representing \textit{all} many-body effects. It is a complex valued quantity whose real part captures the QP energy \textit{renormalization} (i.e., energy shifts) and its imaginary part is proportional to the inverse of the QP lifetime.\cite{mahan2000}

To uncover the role of the solvent environment on the QP excitations,  we \textit{formally} decompose the self-energy into the the term stemming from interaction within the molecule (e.g., solute) and outside (i.e., in the environment), denoted  $\Sigma^m$ and $\Sigma^{env}$. In this \textit{ansatz}, the QP energy of the $j^{\rm th}$ molecular state is
\begin{equation}\label{eq:Eqp}
\begin{split}
\varepsilon_j = \\
&\varepsilon_j^{0} + \bra{\phi_j} \Sigma^m(\omega = \varepsilon_j) +\Sigma^{env}(\omega = \varepsilon_j) - \hat v_{xc} \ket{\phi_j}.
\end{split}
\end{equation}
where $\varepsilon_j^{0}$ is the mean-field eigenvalue (i.e., the $j^{\rm th}$ pole of $G_0$), and $v_{xc}$ is the mean-field exchange-correlation potential (e.g., formulated as a density functional). In practice, the self-energy term is further split into the static and dynamical parts, corresponding to the exchange and the correlation interactions. Here, $\Sigma$ is constructed using the $GW$ approximation, presented in Sections~\ref{sec:Ccalc} and~\ref{sec:Xcalc}, and contributions of the environment are analyzed in the results and discussion section~\ref{sec:resdisc}. We emphasize that the decomposition of the self-energy is formal; we treat the environment and the molecule on the same footing, i.e., no approximations beyond those in $GW$ are applied in this work.

For simplicity, we assume that the molecular states $\ket{\phi_j}$ already represent the Dyson orbitals that diagonalize the many-body Hamiltonian, but our approach can be easily generalized.\footnote{This assumption corresponds to the diagonal approximation to the self-energy. In principle, such a limitation can be lifted, but it holds in common cases.\cite{Kaplan2015,Kaplan2016,Romanova2021offdiag}} Still, the set $\ket{\phi_j}$ does not correspond to the canonical orbitals obtained in the underlying mean-field calculations in the condensed phase. In practice, we first reconstruct $\ket{\phi_j}$ using Wannier functions as detailed in the following section and use them to define the occupied subspace that is sampled by the stochastic formalism described later.

\subsection{Separation subspaces: Wannier Functions}\label{sec:wannierization}

Our goal is to analyze the effect of the environment on the molecular states $\{\ket{\phi_j}\}$ via Eq.~\eqref{eq:Eqp}. Without loss of generality, we will first focus only on charge removal, and $\ket{\phi_j}$ will thus represent an occupied electronic state. We first separate the space spanned by single-particle states into molecular and environmental subspaces. Note that the separation and its construction is rather arbitrary,  but as long as there is a map between the subspace of the isolated molecule and the molecule in the condensed phase, we consider this decomposition as complete. Further, the stochastic approach (discussed in the next subsection) is applicable regardless of choice for the basis vectors; hence, the goal is to provide a practical route to define the molecular subspace. 

Here, we employ the maximally localized Pipek-Mezey Wannier (PMW) functions\cite{Pipek1989,Hoyvik2013,Lehtola2014,Jonsson2017}, $\{\ket{\psi_j}\}$,  obtained from a unitary transformation of the canonical Kohn-Sham (KS) occupied eigenstates. Let  $\{\ket{\psi_j^m}\}$ correspond to states localized on the molecule (i.e., forming the molecular subspace) that are found via the procedure described below. The occupied subspace of the environment is spanned by vectors $\{\ket{\phi^{env}}\}$ obtained through projection:
\begin{equation}\label{eq:environment_states}
    \ket{\phi^{env}_j} = \left(\hat I  - \hat P^m \right) \hat P^{occ}\ket{\phi^c_j}.
\end{equation}
Here $\hat I$ is the identity and $\{\ket{\phi^c_j}\}$ are the \textit{canonical} KS eigenstates of the entire system (i.e., the molecule plus the environment). $\hat{P}^{occ}$ projects onto the occupied subspace, i.e.,  
\begin{equation}\label{eq:Pocc}
\hat{P}^{occ} = \sum_i^{N_{occ}} \ket{\phi^c_i} \bra{\phi^c_i}.
\end{equation}
Here, $N_{occ}$ is the number of occupied states per the simulation cell of the system (including the target molecule). Further, the $\hat P^m$ projector is defined using the PMW states:
\begin{equation}\label{eq:projection_molecule}
     \hat P^m \equiv \sum_{j=1}^{N'_{occ}} \ket{\psi^m_j}\bra{\psi^m_j},
\end{equation}
where $N'_{occ}$ is the number of occupied states on the molecule of interest.

The calculation of the PMW states is iterative and maximizes the following objective function\footnote{We employ the steepest descent algorithm.\cite{Silvestrelli1998,Silvestrelli1999,Thygesen2005}} (see the SI for details)
\begin{equation}\label{eq:objfun}
    \mathcal{P} = \sum_{i}^{N_{occ}}\sum_{\mathcal{A}}^{N_{\mathcal{A}}}[Q^{\mathcal{A}}_{ii}]^{2},
\end{equation}
where $N_{\mathcal{A}}$ represents the number of atoms in the simulation cell. The $Q$ matrix is defined as
\begin{equation}\label{eq:Qmatrix}
Q^{\mathcal{A}}_{ij} = \bra{\psi_{i}} w_\mathcal{A} \ket{\psi_{j}}
\end{equation}
Here, $w_\mathcal{A}$ represents the weight function for each atom type 
$\mathcal{A}$:
\begin{equation}\label{eq:atomweight}
w_{\mathcal{A}}(\br) = \frac{\bar{n}_{\mathcal{A}(\br)}}{\sum_{\mathcal{A}'}^{N_{\mathcal{A}'}}\bar{n}_{\mathcal{A}'}(\br)},
\end{equation}
where $\bar{n}_{\mathcal{A}}(\br)$ is the density function of atom type $\mathcal{A}$ using the simple Gaussian model density.\cite{Oberhofer2009} Finally, $\{\ket{\psi_j}\}$ in Eq.~\eqref{eq:Qmatrix} are
\begin{equation}\label{eq:uni_xform}
\ket{\psi_j} = \sum_{i}^{N_{occ}} W_{ij} \ket{\phi^c_i},
\end{equation}
where, $W$ is the unitary matrix defined as the exponential of an anti-Hermitian matrix (see the SI). The $W$ matrix is iteratively updated, and PMW orbitals correspond to the solutions $\{\ket{\psi_j}\}$ which maximize $\mathcal {P}$.

In practice, we search only for states localized on the molecule of interest. The program with ``locally searching for localized states'' feature is available on Github.\cite{Weng2021} The sums in Eq.~\ref{eq:objfun} thus contain $N_{occ}'$ and $N_{\mathcal{A}}'$ instead of $N_{occ}$ and $N_{\mathcal{A}}$, where the first set of quantities refer to the target molecule. Similarly, the objective function is
\begin{equation}\label{eq:objfun'}
    \mathcal{P'} = \sum_{i}^{N_{occ}'}\sum_{\mathcal{A}}^{N_{\mathcal{A}}'}[Q^{\mathcal{A}}_{ii}]^{2}.
\end{equation}
Note that the unitary transform is still performed on all ($N_{occ}$). However, in each iteration step, $N_{occ}'$ states are identified by having the largest expectation value of the weight function:
\begin{equation}\label{eq:localization}
    L_j = \sum_{\mathcal{A}}^{N_{\mathcal{A}'}} \bra{\psi_j} w_{\mathcal{A}} \ket{\psi_j}.
\end{equation}
The $N_{occ}'$ states with largest $L$ values are used in Eq.~\eqref{eq:objfun'}. The resulting PMW states localized on the target molecule and are denoted as $\{\ket{\psi_j^m\}}$.

Using the set of PMWs, we reconstruct the molecular state $\ket{\phi_j}$ in the condensed phase. \textit{First}, we make an auxiliary calculation for an isolated molecule and obtain its $N_{occ}'$ PMWs, $\ket{\psi^{iso}_j}$. The desired molecular state (corresponding to the KS eigenstate of an isolated system), $\ket{\phi^{iso}_j}$  is represented in the PMW basis:
\begin{equation}\label{eq:phidecomp}
    \ket{\phi^{iso}_j}  = \sum_i^{N_{occ}'} \alpha_{ij} \ket{\psi^{iso}_i}.
\end{equation}
\textit{Next}, we use the molecular PMW basis in the condensed phase, $\ket{\psi^m_l}$, to reconstruct the molecular orbital:
\begin{equation}\label{eq:phiconstr}
     \ket{\phi_j} = \sum_k^{N_{occ}'} \sum_l^{N_{occ}'} \alpha_{ij} \beta_{li} \ket{\psi^{m}_l} ,
\end{equation}
where $\alpha_{ij}$ are taken from the Eq.~\ref{eq:phidecomp} and $\beta_{jk} \equiv \braket{\psi_j^m|\psi^{iso}_k}$. The underlying assumption is that the occupied molecular subspace in the gas phase and condensed phase are the same. In practice, we found one-to-one correspondence between the PMW orbitals and $\beta_{jk} \simeq \delta_{jk}$ and hence, Eq.~\ref{eq:phiconstr} is considerably simplified. The reconstructed molecular orbitals are illustrated in Section~\ref{sec:wannier}.

While we focused on the occupied states, the methodology can be applied to reconstruct the bound empty molecular orbitals. The critical part is the identification of the \textit{bound} unoccupied subspace first for the isolated molecule and then the molecule in the condensed phase. Once done, the reconstruction follows the same steps explained above.

\subsection{Stochastic calculation of the self-energy}\label{sec:stoc}

For calculations of the QP energies, we employ the $GW$ approximation in which the non-local and dynamical $\Sigma(\omega)$ derives from charge density fluctuations (induced by particle removal or addition) that screens the exchange interactions. This approach is typically in good agreement with experiments.\cite{Aryasetiawan1998,Friedrich2006,martin_reining_ceperley_2016} Nevertheless, the $GW$ correlation neglects the quantum fluctuations, which may become important for high energy excitations and/or for unoccupied states.\cite{Vlcek2019,Mejuto-Zaera2021} The method of subspace separations is, however, general and applicable to beyond-$GW$ approaches; it will be explored in the future.  Here, a single-shot perturbative correction is computed within the random phase approximation (RPA) on top of a density functional theory (DFT) starting point (see the SI for details). 

To overcome the high cost of conventional implementations, we employ the recently developed stochastic formalism, which can be applied to extremely large systems owing to its linear scaling.\cite{Neuhauser2014,Vlcek2017,Vlcek2018,Vlcek2018PRM,Vlcek2019,Brooks2020,Weng2020,Romanova2020,Mejuto-Zaera2021} This approach to the one-shot perturbative correction (typically denoted $G_0W_0$) seeks the QP energy by randomly sampling the single-particle states and decomposing the operators in the real-time domain. The stochastic formalism has been described in detail in previous work\cite{Neuhauser2014,Vlcek2017,Vlcek2018} and it is briefly revisited in the SI. In the following two subsections, we describe the connection between random sampling and the decomposition of the correlation and exchange self-energies.

\subsubsection{Correlation contribution}\label{sec:Ccalc}

As stated above, the correlation part of the $GW$ self-energy, $\Sigma_c$, is governed by the charge density fluctuations. Recently, we have formulated a stochastic approach to decompose $\Sigma_c$ (Ref.~\onlinecite{Romanova2020}), which is also applied to the calculations of the $\Sigma_c^{m}$ and $\Sigma_c^{env}$ in this work. Here, $m$ and $env$ denote the molecule and the environment. 

Within the stochastic formalism, we employ random states $\zeta$ to decompose the GF. The resulting expression for the correlation self-energy $\Sigma_c(t)$ explicitly depends on the induced charge density potential $u_{\bar{\zeta}}(\br,t)$
\begin{equation}
  \bra{\phi_j} \Sigma_c(t) \ket{\phi_j} \simeq \frac{1}{N_{\zeta}} \sum_{\zeta} \int \phi(\br) \zeta(\br,t) u_{\bar{\zeta}}(\br,t) \dd \br,
\end{equation}
where $u_{\bar{\zeta}}(\br,t)$ is in practice computed from the retarded response potential due to the  $\delta n(t)$:
\begin{equation}\label{eq:ru}
\tilde u_{\zeta}(\br,t) = \int \nu(\br,\bp) \delta n(\bp,t) {\dd}\br' .
\end{equation}
The definitions of $u_{\bar{\zeta}}(\br,t)$ and $\tilde u_{\zeta}(\br,t)$ are provided in the SI.

Next, we separate the induced time-dependent charge density $\delta n(\br,t) = n(\br,t) - n(\br,0)$ into two terms: for the time-dependent density of the molecule, i.e., solute, and the environment, i.e., solvent (distinguished by superscripts):
\begin{equation}\label{eq:nt}
    n(\br,t) = n^m(\br,t) + n^{env}(\br,t).
\end{equation}
Both densities are expressed via the linear combination of PMW states (Section~\ref{sec:wannierization}).  In our stochastic approach, the time-dependent $n^{m}$ and $n^{env}$ are sampled using a set of random vectors $\{ \eta^k \}$ confined to the corresponding (and mutually orthogonal) subspaces $k$ ($k=m, env$):
\begin{equation}
n^k(\br,t)=\frac{1}{N_{\eta}}\sum_{l}^{N_{\eta}}|\eta^k_l(\br,t)|^2.
\label{eq:TDdensity_env}
\end{equation}
Here, $N_{\eta}$ represents the number of random vectors used in the sampling (we employ $N_\eta = 16$; see the SI for details). 
The random vector is prepared by projections:
\begin{equation}\label{eq:etak}
\ket{\eta^k} = \hat{P}^k \hat{P}^{occ} \ket{\eta},
\end{equation}
where the case of $\hat{P}^m$ is given by Eq.~\ref{eq:projection_molecule} and $\hat{P}^{env} = \hat I - \hat{P}^m$. The projector $\hat P^{occ}$ is defined by Eq.~\eqref{eq:Pocc}.

The time propagated random vector $\ket{\eta^k(t)}$ in Eq.~\ref{eq:TDdensity_env} is:
\begin{equation}\label{eq:etat}
    \ket{\eta^k(t)} = U_{0,t}[n(t)] \ket{\eta^k},
\end{equation}
where $U_{0,t}$ is the time evolution operator defined as
\begin{equation}\label{eq:toe}
    U_{0,t}[n(t)] = e^{-iH_0[n^m,n^{env}]t}.
\end{equation}
Here, $H_0$ is the mean-field Hamiltonian that is a functional of the total time-dependent charge density (i.e., containing contributions of both subspaces). Propagation using the full charge density ensures that the correlated dynamics between the two subsystems is captured. 

\subsubsection{Exchange contribution}\label{sec:Xcalc}
The exchange self-energy is clearly defined via the non-local integral involving the density matrix, $\rho$, and the bare Coulomb interaction,  $\nu$:
\begin{equation}\label{exc}
    \Sigma_x = \iint  {\varphi_j}^*(\br) \nu(\br,\br') \rho(\br,\br') \varphi_j(\br') \dd{\br'} \dd{\br},
\end{equation}
where $\varphi_j$ is a single-particle state. The density matrix is 
\begin{equation}\label{eq:dm}
   \rho(\br,\br') = \sum_i^{N_{occ}'} {\varphi_i}(\br) \varphi^*_i(\br').
\end{equation}

In practice, we disentangle the contributions from various parts of the system similar to the procedure applied to the correlation self-energy. Due to the orthogonality of the molecular and environment states, the density matrix is simply decomposed into the corresponding contributions $\rho^m$ and $\rho^{env}$. The former is analogous to the solution for an isolated molecule. The contribution of the environment is employs states $\{\ket{\phi_j^{env}} \}$ obtained by the projection in Eq.~\eqref{eq:environment_states}.

\section{Results and Discussions}\label{sec:resdisc}

We now demonstrate the methodology and investigate the role of solute-solvent interactions on the PES of phenol, thymine, and phenylalanine in water. The molecules are selected based on their relevance in chemistry and (variable) structural complexity. Phenol in water is a ubiquitous motif in biological chromophores, and hence this system has attracted interest in the past.\cite{Ghosh2012,Riley2018} Thymine and phenylalanine are studied in the context of DNA molecules\cite{Close2004,Crespo-Hernandez2004,Slavicek2009,Ghosh2011,Chakraborty2017} and amino acids.\cite{Lee2003,Papp2012,Close2011,Sadybekov2017,Barrozo2019} While phenol is the simplest and represents a ``rigid'' system (only a hydroxyl group is attached to the benzene ring), thymine has a methyl group attached to the ring, which can freely rotate. Phenylalanine has a much longer amino-acid group attached to the benzene ring. The molecular structures are provided in Fig.~\ref{fig:mol_structures}.

The system geometries are obtained using Molecular Dynamics simulations (see the SI). From the trajectory of the calculation, we extract five uncorrelated solute-solvent structures and perform the stochastic many-body calculations. Ground state DFT is used to obtain the starting point for $GW$: the calculations are performed on a regular real-space grid with generalized gradient approximation to exchange and correlation in combination with the Troullier Martins norm-conserving pseudopotentials.\cite{Troullier1993} The computations for isolated molecules are denoted as $iso$ henceforth; the full QP spectra are computed for the solvated systems (see the discussion below) captured by simulation cells with the periodic boundary condition with up to 1056 electrons. The stochastic calculations of QP energies are converged with respect to the number of stochastic samples such that the errors are $\le 0.07$ eV for the frontier state. The random sampling is efficient -- to reach this level of statistical error, the simulation took only 438, 988, and 1315~CPU$\cdot$hrs on a cluster computer equipped with CPUs a clock speed of 2.40GHz. 

In the rest of this section, we first discuss the orbital reconstruction, followed by the analysis of the QP renormalization in the liquid phase. Finally, we show the full theoretical PES.

\subsection{One-to-one correspondence of occupied subspace}\label{sec:wannier}

\begin{figure}
  \includegraphics[width=\linewidth]{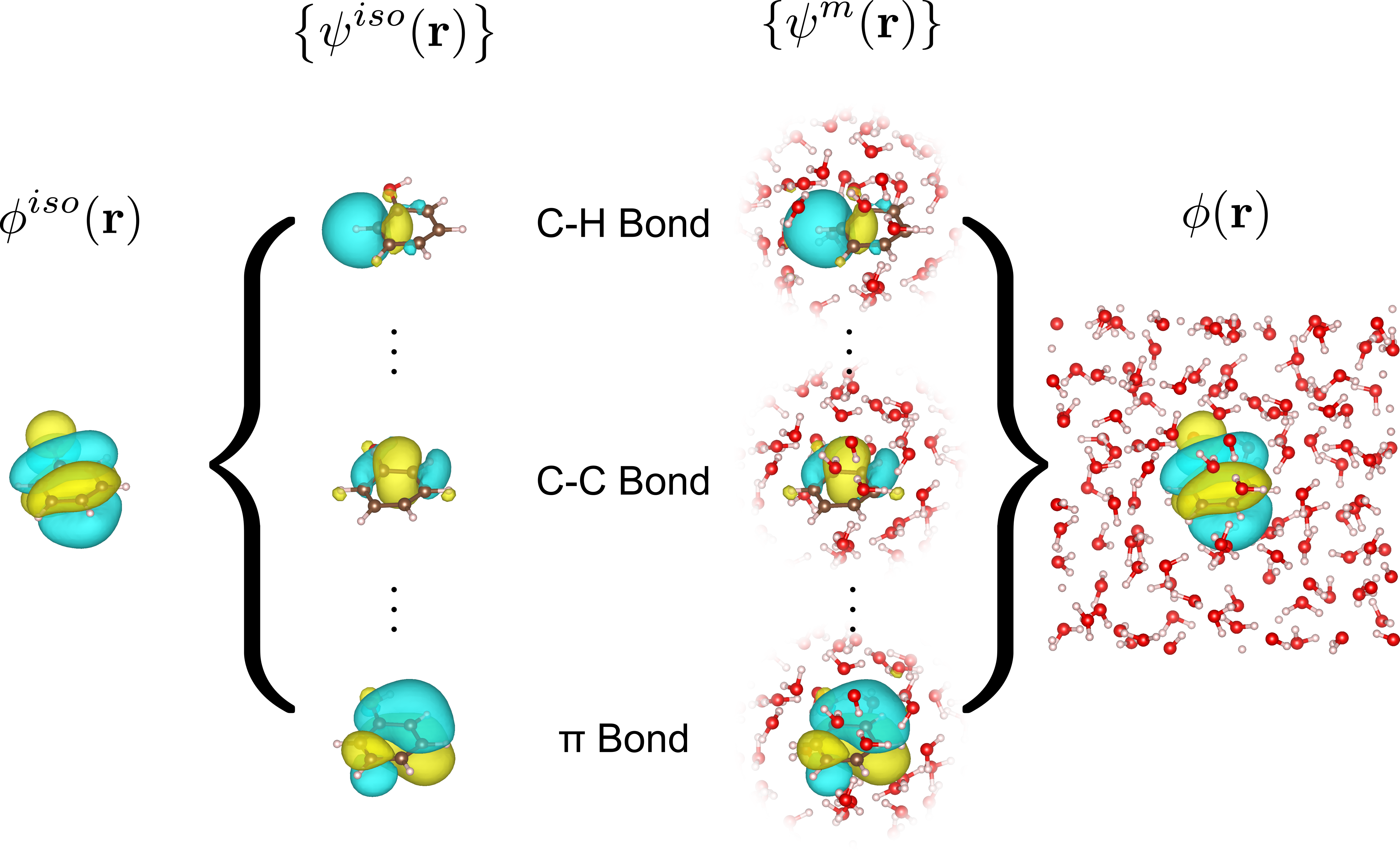}
  \caption{Three typical PMW functions of isolated and solvated phenol molecule, respectively (Middle). The PMW functions are labelled by chemical bonds due to their spatial distribution. The eigenstate, e.g., HOMO, of the isolated molecule is expressed as a linear combination of $\ket{\psi^{iso}_j}$ (Left). Using the same set of coefficients, the molecular state of the solvated molecule is constructed as a linear combination of $\ket{\psi^m_j}$ (Right).}
  \label{fig:orbcon}
\end{figure}

Following the ground state DFT calculations, we determine the PMW states for both the isolated and solvated solute molecules. Fig.~\ref{fig:orbcon}, shows the typical PMW functions for phenol molecule. In the rest of this paper, we consider only the occupied subspace; a similar approach is, however, applicable to empty states as discussed in Section~\ref{sec:wannierization}.  

The PMW functions have highly localized spatial distribution, and it is appealing to interpret them as bonding orbitals that characterize the entire occupied subspace. In this (convenient) framework, one can expect that PMW functions of the isolated and solvated systems are related. Indeed, our calculations indicate that we can directly map the isolated and solvated system PMW functions onto each other in all the cases studied (Fig.~\ref{fig:1to1_phenol}). The spatial overlap between the corresponding PMW functions is at least 99\% for all the three molecules (Table~\ref{tab:PMWoverlaps}). 

The correspondence between PMWs of isolated and embedded systems is likely common, and PMWs can thus be leveraged to reconstruct the molecular orbitals. As discussed in section~\ref{sec:wannierization}, it is not, however, necessary that there is a one-to-one map. Instead, the only underlying assumption is that the selected subspace is the same for both the solvated and isolated molecules. In particular, this is one of the advantages of the stochastic methods, which merely require the knowledge of the molecular \textit{subspace} (not the states).

Fig.~\ref{fig:orbcon} shows the reconstructed highest occupied molecular orbital (HOMO) for the phenol in water (the right panel) versus the counterpart of the isolated one (the left panel). The reconstructed $\ket{\phi}$ enters Eq.~\eqref{eq:Eqp}, and it is used to evaluate the role of the environmental effects. 

\subsection{HOMO QP energies and energy shifts}\label{sec:homo}

\begin{figure*}
  \includegraphics[width=0.6\linewidth]{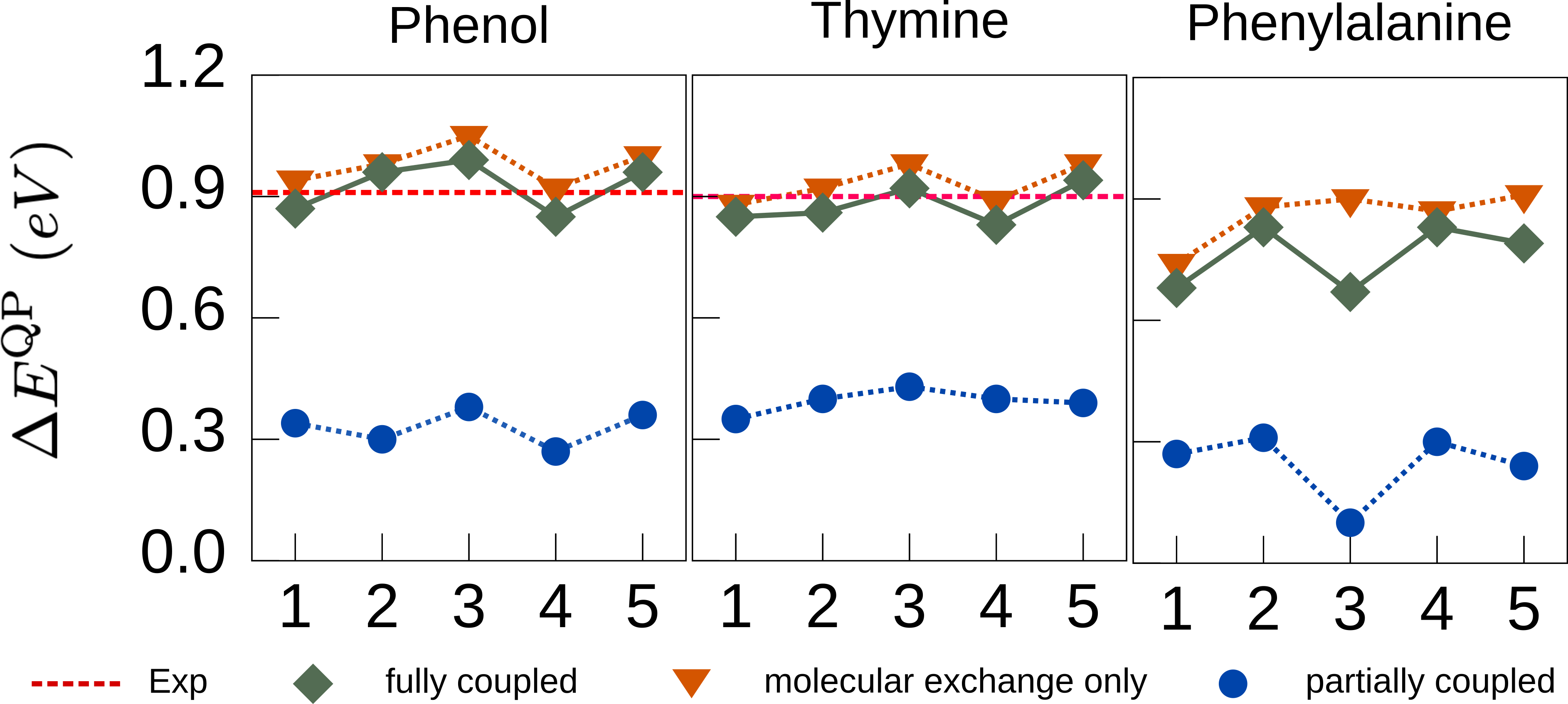}
  \caption{HOMO QP energy shifts with respect to the snapshots along the MD trajectory of the three solvated molecules. The red dashed lines represent the experimental results for phenol\cite{Ghosh2012} and quantum chemistry calculations for thymine\cite{Ghosh2011}.}
  \label{fig:QPES_homo}
\end{figure*}

The QP energies of frontier states are of primary interest in the charge transfer process as they correspond to the negative of the first ionization/detachment energy. We employ the reconstructed orbitals discussed above. The solvent affects the QP energy (Eq.~\eqref{eq:Eqp}) via (i) variation of the molecular (solute) geometry and (ii) changes in the total self-energy and presence of the $\Sigma^{env}$ term for the solvated molecule.

We first quantify the effects of the solute structure as changes of the QP energy purely due to the geometry variation in the absence of the environment. In practice, we compare the average of the QP energies for isolated molecules (with geometries extracted from the MD trajectory) against the results for the system in its ground state geometry in the gas phase (see Table~\ref{tab:QPE_iso}). On average, the absolute differences between the two are merely 0.05, 0.22, and 0.15~eV for phenol, thymine, and phenylalanine, respectively. Surprisingly, the $GW$ values appear to be less sensitive to the structural changes than the underlying DFT (Table~\ref{tab:PBE_iso}). More importantly, the $GW$ method provides results that are in good agreement with experimental gas-phase IP measurements (albeit with a systematic shift, see Table~\ref{tab:QPE_iso}).

Although the HOMO QP energies do not deviate significantly from the gas-phase values  \textit{on average}, they can markedly differ for a particular configuration (Fig.~\ref{fig:Eqp_all}). The standard deviations ($\sigma$) stemming from structural variations of isolated molecules are 0.08, 0.11, and 0.28 eV for phenol, thymine, and phenylalanine (Table~\ref{tab:QPE_iso}), respectively. Clearly, the spread of the results (captured by $\sigma$) increases with the molecular complexity  (i.e., from phenol to phenylalanine). Furthermore, the structural changes are associated with electronic state reordering, i.e., the QP energies associated with particular states are (de)stabilizing various states differently  and  QP HOMO is not identical to the DFT HOMO orbital. Hence, the non-local dynamical correlations in $\Sigma^m$ are responsible for significant energy renormalization. For instance, the HOMO orbital is not the same as predicted by DFT in the majority of the calculations for phenylalanine (see Table~\ref{tab:reodering} for details). 

In the remaining part, we focus on the QP energy renormalization stemming purely from the electron-electron interactions due to the water solvent. Again, we quantify them as QP energy shifts relative to the QP energy of isolated molecules. The results are presented in Fig.~\ref{fig:QPES_homo} (the dark green solid lines) and Table~\ref{tab:QPEshifts}, where the averaged QP energy shifts over 5 snapshots are $+0.92\pm0.06$, $+0.88\pm 0.05$, and $+0.76\pm0.08$~eV for phenol, thymine, and phenylalanine, respectively. The QP energies are significantly destabilized, and the ionization potential thus decreases. The predicted QP energy`` shifts are in excellent agreement with previous PES measurements and quantum chemistry calculations.

The environment affects the QPs more than by just rigidly shifting the excitation energies. Indeed, the QP energy variation for phenol and phenylalanine in water increases by 75\% and 18\% (to 0.14 and 0.33 eV). In contrast, $\sigma$ for thymine in water slightly decreases by 11\%. The numerical results are provided in Table~\ref{tab:QPE_sol}. This indicates that the many-body interactions between the solute and solvent can strongly depend on the actual molecular arrangements, and the effects are further analyzed in the following subsection.

\subsection{``Fully coupled'' correlation vs.~``partially coupled'' correlation}
\label{section:correlation}

\begin{figure*}
  \includegraphics[width=0.6\linewidth]{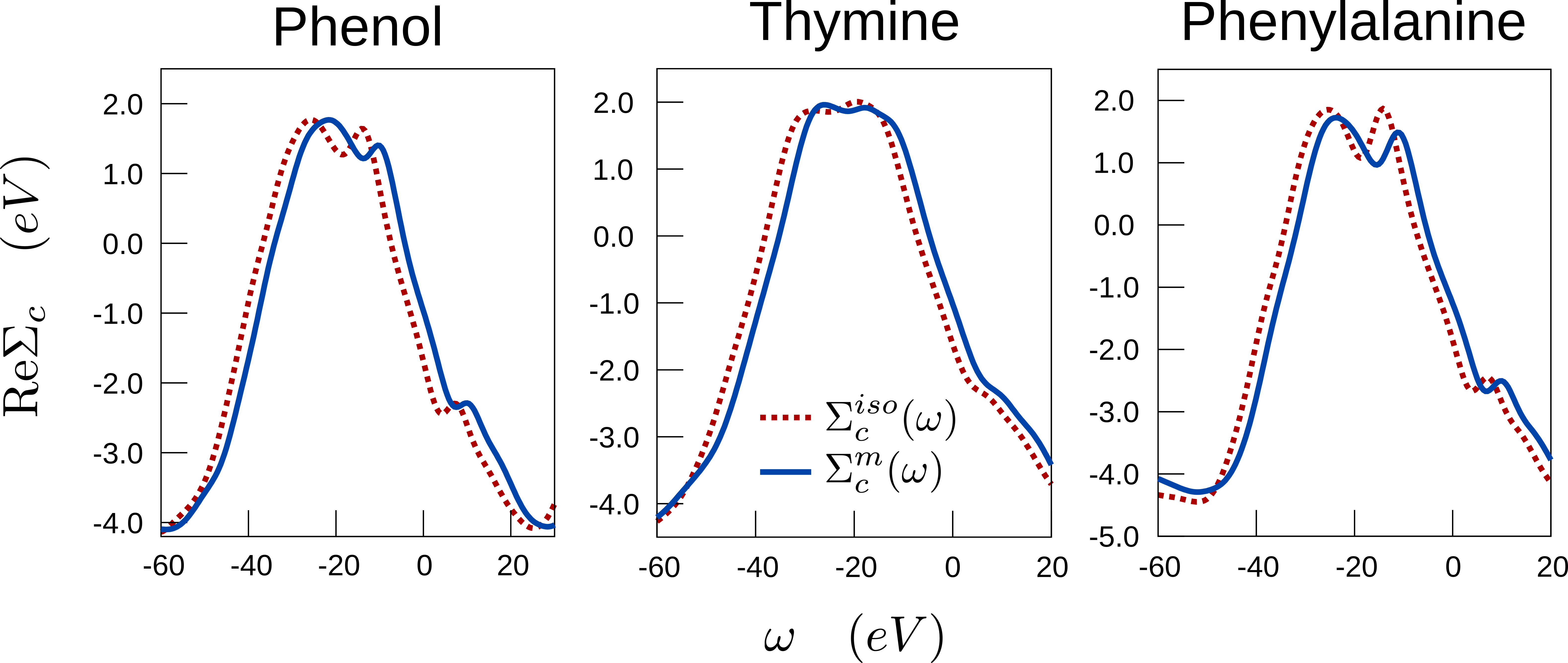}
  \caption{Comparisons of the real-part of the correlation self-energy with (solid line) and without (dotted line) the solvent-to-solute feedback for the three molecules. When the solute-to-solvent feedback is accounted, the self-energy curves derived from the charge density fluctuations on the solvated molecules ($\delta n^m$) are unanimously shifted to lower frequency region with respect to that of the isolated molecules.}
  \label{fig:RESE}
\end{figure*}

To understand the changes in the QP energies, we decompose the self-energy into exchange and correlation terms stemming from the molecule and the environment. The exchange interaction generally leads to the QP stabilization, and it is governed by solute molecular states. The contribution of $\Sigma_x$ due to the solvent is small ($\le 0.16$ eV, Table~\ref{tab:env_exc}) due to the limited spatial overlap between the the molecular orbitals and the remainder of the system (Fig.~\ref{fig:QPES_homo}, the dark orange dotted line and Table~\ref{tab:env_exc}). In practice, the non-local and dynamical correlation self-energy thus captures the key aspects of mutual coupling between the solute and solvent. 

To understand the effect of the environment on $\Sigma_c$, we will, for simplicity, consider the first ionization energies, i.e., HOMO states. First, we note that the relative contribution of environment correlation $\Sigma_{c}^{env}$ (i.e., the fraction of the total $\Sigma_{c}$) ranges between 30\% and 47\% (Table~\ref{tab:Cenv_percent}) and the remaining portion, $\Sigma_{c}^{m}$, is due to the dynamical interactions of electrons localized in the molecule. 

The environmental effects are, however, not represented solely by $\Sigma^{env}$. In fact, the charge removal from a molecular state, $\ket{\phi_j}$, leads to dynamical interactions with electrons in the surrounding water molecules. Besides the ``direct'' effect (i.e., solute-to-solvent), the induced charge density fluctuations in water also create a solvent-to-solute feedback mechanism. Hence, the induced charge density in the environment leads to secondary fluctuations in the molecule and so on. This coupling is inherently captured by the charge density in each subsystem, $n^k(\br,t)$, which is subject to the time evolution governed by the \textit{total} charge density (Eq.~\eqref{eq:toe}).  In this case, the real-time formalism is particularly appealing as it enables capturing strong (i.e., even non-linear) couplings.\cite{Runge1984,Cossi2001,Marques2004,Adamo2013}

We now compare the self-energies of the molecule in the environment, $\Sigma_{c}^{m}$, with an isolated one,  $\Sigma_{c}^{iso}$,  i.e., without the environmental feedback, .\footnote{The calculations for the isolated molecule make use of the geometry extracted from the MD trajectory of the solvated system} The comparison provides an estimate of the solvent-to-solute coupling magnitude. The real parts of the HOMO self-energy $\Sigma_{c}^{iso}$ and $\Sigma_{c}^{m}$ are shown in Fig.~\ref{fig:RESE} for one of the snapshots along the MD trajectory. The observed shift with respect to each other is due to the solvent. In practice, the spectral features are shifted closer to QP energy and generally enhance the correlation energy contribution. The changes do not correspond merely to a rigid shift of the self-energy and distinct, i.e., various maxima of the curve are affected differently. 

If we recompute the QP energy by combining  $\Sigma_{c}^{env}$ with $\Sigma_{c}^{iso}$, the correlations contain only screening stemming either from the molecule or the environment but the indirect solvent-to-solute contributions are missing. To distinguish this result, we label this QP energy as ``partially-coupled''. Fig.~\ref{fig:QPES_homo} shows that in this case the QP energy shifts decrease by as much as 68\% (to 0.33, 0.39, and 0.24 eV for phenol, thymine, and phenylalanine, respectively, Table~\ref{tab:QES_SSB}). Hence, the induced interactions between the charge densities represent a significant component of the QP renormalization. This observation illustrates the need for a fully coupled description of the molecule and the environment.

Note that the approach proposed here is distinct from the previous calculations employing $GW$ within the QM/MM embedding,\cite{Li2016,Li2018} employing reaction field describing the solvent, which is often based on the classical charge response model.\cite{Tsiper2001} This embedding accounts for the environment response in the screened Coulomb interaction term ($W$) through the inclusion of the solvent polarizability. In practice, the correlation term contains contributions from the induced charge density on the molecule and the environment, but their dynamical coupling (i.e., secondary solvent-to-solute interaction) is neglected except for structural changes of the solute molecule and its orbitals. In contrast, the approach proposed here treats both subspaces consistently and fully coupled (within the $GW$ approximation).
Further, the environmental effects are present in all the terms entering the self-energy evaluation, i.e., not only  $W$. This is particularly important for calculations that include higher-level treatment beyond $GW$.\cite{Vlcek2019,Mejuto-Zaera2021}

\subsection{Simulated Photo-emission Spectra}\label{sec:PES}

\begin{figure*} 
  \includegraphics[width=\linewidth]{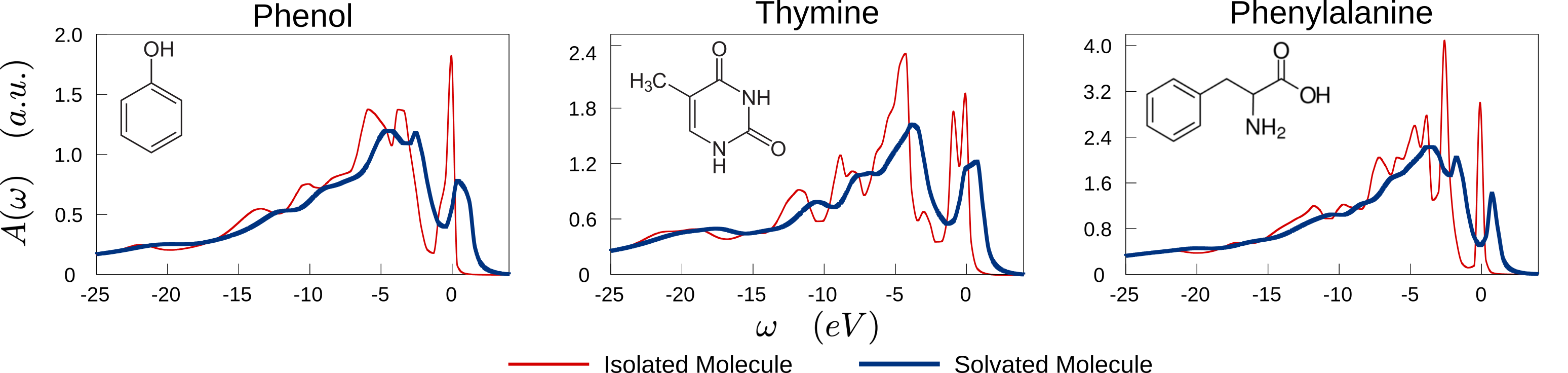}
  \caption{Simulated photo-emission spectra of the three molecules depicted in the insets. The spectral functions for an isolated molecule and its solvated counterpart are distinguished by color.}
  \label{fig:spectra_sum} 
\end{figure*}

In the final step, we turn to the prediction of the entire photo-emission spectra of the solvated molecules represented by the spectral function $A(\omega)$ (Section~\ref{sec:QPandSE}). Here, the total  $A(\omega)$ curve is a sum over the imaginary components of the GFs from each valence state.\footnote{Only one selected system snapshot along the MD trajectory is considered in this section.}  
Fig.~\ref{fig:spectra_sum} shows a comparison between $A(\omega)$ for the isolated and solvated molecules. The zero on the frequency (energy) axis is aligned with the HOMO QP energy of the isolated system. The common feature is the destabilization of the molecular states, i.e., shifts to higher energies. 

For phenol, the profiles of the PES in two phases are similar, although the height of the frontier state's peak is significantly reduced and simultaneously broadened. This indicates a strong QP lifetime decrease due to the electron-electron interactions with the water solvent. Note that for the spectra of isolated molecules, the peak width is artificially widened by the finite time evolution employed when computing $\Sigma_c^{iso}$; since this finite broadening affects the spectrum uniformly, we consider only the relative changes in the peak width with respect to the HOMO state. Further, note that we consider only the \textit{electron-electron scattering}, but, in practice, additional broadening will be observed due to vibrational couplings. The results presented in Fig.~\ref{fig:spectra_single} clearly indicate that even if only electronic degrees of freedom are considered, the QP lifetime shortening is significant for all valence states and the spectral broadening progresses with the hole energy (i.e., the deep valence holes correspond to wide peaks).

For thymine, besides the energy shift, the overall spectrum is much smoother as molecular distortions and solvent couplings lift the QP state degeneracy. In general, the peak widths of the solvated systems are on average $50\%$ increased (see also Table~\ref{tab:ImSE}). These effects are most obvious in the PES of phenylalanine: the spectral function is destabilized, the QP lifetime decreases, and we observe almost complete smearing of the entire spectrum below the HOMO level.  The QP lifetime of phenylalanine's frontier state is shortened most significantly, by as much as $90\%$,  leading to a drastically reduced peak height in Fig.~\ref{fig:spectra_sum}.

For completeness, we analyze the energy shifts $\Delta E^{QP}$ of all valence states (see    Tables~\ref{tab:QPphenol}-\ref{tab:QPphenylalanine}). For phenol and phenylalanine, $\Delta E^{QP}$ is practically constant (on average 0.94 and 0.86 eV). For thymine, the bottom valence states are more destabilized compared to the top valence region; yet the average shift for each state is comparable to the other two molecules  (0.87~eV). For a specific system configuration, the environmental effect is consistent for all valence states.

\section{Conclusion}

In summary, a fully ab-initio stochastic many-body method is established to calculate QP energy for molecules in the condensed phase. Here, the separation between the electronic states is general, and the methodology treats the entire system at the same level of theory. We adopt the $GW$ approximation to describe the QP excitations. The approach combines the random sampling of operators in mutually orthogonal subspaces corresponding to the molecule and the environment. We presented a practical route to reconstruct the molecular subspace in liquid water via a linear combination of localized PMW functions. Owing to the linear scaling of the stochastic methodology, realistic calculations for systems with thousands of electrons are thus possible.

The approach is tested on three solute-solvent combinations. By separating the interactions, we find that the environment correlation energy $\Sigma_c^{env}$ accounts for $\sim40\%$ of the QP renormalization. Further, the solvent effects are responsible for dynamical coupling to the correlations within the molecular subspace and contribute by up to 68\% of the overall energy shift. The simulated PES of solutes shows the following characteristic features: destabilization (i.e., red shift) of the peaks, smearing of the entire spectrum, and significant QP lifetime shortening.

The methodology is general and can be applied to liquid and solid phases. It provides direct access to the charge excitation energies as well as the photo-emission spectra and their changes due to the interactions with the environment. Thus, this approach provides a new perspective on ionization and charge transfer processes, and ultimately the chemical reactivity and opto-electronic properties of molecules in complex systems. 

\section*{Supplementary Material}
The supplementary material provides the supporting information for the details of DFT and GW calculations, Pipek-Mezey wannier functions, and spectral functions. Supplementary tables and figures indicated in the texts are also provided in this document.

\begin{acknowledgements}
The authors want to acknowledge Prof.~Thuc-Quyen Nguyen for fruitful discussions. This work was supported by the NSF CAREER award through Grant No. DMR-1945098. The calculations were performed as part of the XSEDE computational Project No. TG-CHE180051. Use was made of computational facilities purchased with funds from the National Science Foundation (CNS-1725797) and administered by the Center for Scientific Computing (CSC). The CSC is supported by the California NanoSystems Institute and the Materials Research Science and Engineering Center (MRSEC; NSF DMR 1720256) at UC Santa Barbara.
\end{acknowledgements}

\section*{DATA AVAILABILITY}
The data that support the findings of this study are available from the corresponding author upon reasonable request.

\bibliography{Wannier_Condensed}

\end{document}